\documentclass[prb,aps,amsmath,showpacs,preprint]{revtex4}
\usepackage{graphicx}% Include figure files

\usepackage{dcolumn}% Align table columns on decimal point

\usepackage{bm}% bold math

\begin{document}
%\begin{center}
\title{First principles study of rare-earth oxides} 
%\vspace{0.2cm}
\author{L. Petit$^{1,2}$, A. Svane$^{2}$, Z. Szotek$^{3}$, and W.M. Temmerman$^{3}$}
%\vspace{0.2cm}
\affiliation{
$^{1}$ {\it Computer Science and Mathematics Division, and Center for Computational} \\
{\it Sciences, Oak Ridge National Laboratory, Oak Ridge, TN 37831, USA} \\
$^{2}$ {\it Department of Physics and Astronomy, University of Aarhus,} \\
{\it DK-8000 Aarhus C, Denmark} \\
$^{3}$ {\it Daresbury Laboratory, Daresbury, Warrington WA4 4AD, UK} }
%\end{center}

\date{\today}

\begin{abstract}
The self-interaction-corrected local-spin-density approximation is used 
to describe the electronic structure of dioxides, REO$_2$, and sesquioxides,
RE$_2$O$_3$, for the rare earths, RE=Ce, Pr, Nd, Pm, Sm, Eu, Gd, Tb, Dy and Ho. The valencies 
of the rare earth ions are determined from total energy minimization.
We find Ce, Pr, Tb in their dioxides to have the tetravalent configuration, while
for all the sesquioxides 
the trivalent groundstate configuration is found to be the most favourable. 
The calculated lattice constants for these valency configurations are in good agreement with experiment. 
Total energy considerations are exploited to show the link between oxidation and $f$-electron delocalization, 
and explain why, among the dioxides, only the CeO$_2$, PrO$_2$, and TbO$_2$ exist in nature. 
Tetravalent NdO$_2$ is predicted to exist as a metastable phase -
unstable towards the formation of hexagonal Nd$_2$O$_3$.

\end{abstract}

%\pacs{}
%\narrowtext
%\twocolumn
 
\maketitle

\section{Introduction}
 
Even though the rare-earth (RE) elements readily oxidize, they do so with varying strength.\cite{moritz} Ce metal
oxidizes completely to CeO$_2$ in the presence of air. Pr occurs naturally as Pr$_6$O$_{11}$,
exhibiting a slightly oxygen deficient fluorite structure. The stoichiometric fluorite structure
PrO$_2$ exists under positive oxygen pressure. The rare-earth  oxides from Nd onwards, 
with the exception of Tb, all occur naturally as sesquioxides, RE$_2$O$_3$.
 Tb oxide occurs naturally as Tb$_4$O$_7$,
and transforms into TbO$_2$ under positive oxygen pressure.

The present study is concerned with the valencies of the oxides of 
Ce, Pr, Nd, Pm, Sm, Eu, Gd, Tb, Dy and Ho. Under suitable
conditions, all the RE elements form a sesquioxide,~\cite{eyring} 
and there is general agreement that, in the corresponding groundstate, the
rare-earth atoms are in the trivalent, RE$^{3+}$ configuration.~\cite{tanaka,moewes} 
%The stoichiometry of the sesquioxides can be understood in terms of a trivalent rare-earth ion configuration,
Each rare-earth atom donates 3 electrons to the strongly electronegative O ions, and the remaining 4$f$
electrons stay strongly localized at the rare-earth site. In the lighter lanthanides, the $f$-electrons
are less tightly bound, resulting in compounds that display a larger oxygen coordination number,
CeO$_2$ and PrO$_2$.
Similarly in Tb, the extra electron on top of the half filled
shell is less tightly bound resulting in a valency larger than 3+.

Only Ce, Pr, and Tb
form dioxides, and with respect to the valency of the corresponding RE-ions the debate is still ongoing.
Two conflicting points of view, both based on the interpretation of core-level spectroscopy studies,
describe the dioxide groundstate as either tetravalent~\cite{karnatak,hanyu,wuilloud} 
or intermediate-valent.~\cite{bianconi,ogasawara,butorin} The intermediate-valent 
interpretation uses the spectroscopic data
to obtain the model parameters entering the Anderson impurity Hamiltonian.~\cite{anderson} 
By solving this
many-body Hamiltonian~\cite{gunnarson} 
the initial state is derived from the final-state photoemission spectra. 
It should be noted however that Wuilloud {\it et al.}~\cite{wuilloud} 
derive the tetravalent goundstate for CeO$_2$,
using the same method. Marabelli and Wachter~\cite{marabelli,wachter} conclude from their optical
spectroscopy measurements that CeO$_2$ is a tetravalent insulator. 
Neutron scattering results on PrO$_2$, by Kern {\it et al.}~\cite{kern} 
were interpreted in terms of a localized 4$f^1$ groundstate configuration of the Pr ion.
Similarly,
recent neutron spectroscopy measurements of the magnetic exitations in PrO$_2$ 
have also been interpreted in terms of a tetravalent 4$f^1$ configuration.~\cite{boothroyd}
With respect to the XPS results on PrO$_2$, the proponents of the tetravalent picture explain
the data in terms of the coexistence of localized and delocalized $f$-electrons.~\cite{karnatak}

The rare-earth oxides find important applications in the catalysis, 
lighting and electronics industries and thus
are extremely interesting compounds in their own right. A further reason for the strong
interest in the RE-dioxides is their relation to the 123-cuprates REBa$_2$Cu$_3$O$_7$. 
One notices for example that it is again  Ce, Pr and Tb that distinguish themselves 
from the other rare earth elements, 
as their corresponding cuprates do not support superconductivity.~\cite{chryssikos} Also in this case, no
definite picture of the valency of the rare-earth ion has yet emerged.
Similar to the intermediate-valency 
interpretation of the PrO$_2$ data, it has been suggested that the
suppression of superconductivity in PrBa$_2$Cu$_3$O$_7$, is associated with the existence of intermediate-valent
Pr ions.~\cite{fehrenbacher} The determination of the valency configuration in the dioxides might
thus be very useful for the elucidation of the mechanism behind superconductivity in the cuprates. 

Any theoretical description of the rare-earth oxides needs to take into account the strong on-site
$f$-$f$ interactions that push the 4$f$-electrons towards localization. 
Using the Anderson impurity Hamiltonian, a possible approach is to derive 
the intraatomic Coulomb energy U from experiment, as was
done in connection with the interpretation of XPS data.\cite{karnatak} There is also a possibility
of obtaining the parameters of the Hamiltonian from first principles calculations, as was done
by McMahan {\it et al.}.~\cite{mcmahan} 
Electronic structure calculations, based on density functional theory have been very 
successful in describing the cohesive properties of solid-state systems with itinerant valence electrons.
However the exchange and correlation effects of the homogeneous electron gas,
which are included in the local-spin-density (LSD) approximation, are insufficient to account for
the strong correlations experienced by the localized 4$f$-electrons in the rare-earth
compounds. 
With respect to the rare-earth oxides, the band structure calculations 
that we could find in literature are for CeO$_2$,~\cite{koelling,skorodumova,fabris} 
PrO$_2$,~\cite{koelling,dabrowski} Ce$_2$O$_3$,~\cite{skorodumova,fabris}, and RE-sesquioxides in general.~\cite{hirosaki}
From these studies it becomes clear that for example 
CeO$_2$ is best described in terms of itinerant $f$-electrons, 
whilst for Ce$_2$O$_3$, considering one Ce $f$-electron as part of the core leads to better agreement with the
experimental lattice parameter.~\cite{skorodumova} In this case, the $f$-electron groundstate 
configuration, and consequently the choice of calculational tool, is determined from a comparision to 
empirical data. Fabris {\it et al.}~\cite{fabris} show that the LDA+U approach gives 
good agreement with experiment for both Ce polymorphs, but the method requires input of a Hubbard 
U parameter (3 eV for Ce). In PrO$_2$ the LSD 
description reveals a large $f$-peak at the Fermi
level, in disagreement with the fact that PrO$_2$ is an insulator. 

In the self-interaction-corrected (SIC) local-spin-density approach~\cite{thesic2,thesic1} 
of the present paper, both localized and delocalized $f$-electrons are assumed to coexist. 
The delocalized $f$-levels 
move in the LSD potential, their exchange and correlation energies are those derived on the basis
of the homogeneous electron gas,
and their ability to form bands leads to a gain in
hybridization energy. 
The localized $f$-levels acquire core-like character by correction of the LSD
total energy functional for their spurious self-Coulomb and self-exchange-correlation energies.~\cite{perdew} 
This leads to an additional negative potential term and effectively prevents any hybridization.
In the SIC-LSD
approach, the localized and itinerant $f$-states are treated on an equal footing, and
configurations with varying numbers of localized $f$-states can be compared with respect to their
total energies. Consequently the preferred groundstate configuration, as far as the number of
localized $f$-electrons is concerned, can be determined from the global total energy minimum. Namely by
comparing the total energies resulting from different electronic configurations, one can determine
whether it is favourable for the $f$-electrons to localize or to contribute to the band formation. 
The method has previously been succesfully applied to describe 
the valencies of the RE elements~\cite{nature} and compounds (see for example
[\onlinecite{review}] and references therein). 

The remainder of this paper is organized as follows. In section II and III respectively the results for the
RE-dioxides and RE-sesquioxides are presented and compared to experiment. 
In section IV, the oxidation/reduction process is studied, and in section V, we draw our conclusions.

\section{The RE-dioxides}

The SIC-LSD approach has been implemented using the tight-binding linear muffin-tin orbital (LMTO)
method~\cite{oka} in the atomic sphere approximation (ASA). The spin-orbit interaction is included in the
Hamiltonian. Empty spheres are inserted on high symmetry interstitial sites. 
 The valence panel includes the 6s,
5p, 5d and 4f orbitals on the rare-earth atom, and the 2s and 2p on the O atom, while the higher order
$\ell$ orbitals of O ($3d$ and $4f$), as well as all degrees of freedom of the empty spheres are treated as
downfolded.\cite{oka} A separate energy panel is used to describe the semicore $5s$ states of the rare earth.
In the valence band, we need to distinguish between those $f$-electrons that are 
SIC-localized, and the delocalized $f$-electrons, which together with the $s$, $p$ and $d$ electrons
constitute the valence electrons. Different electronic configurations will give rise to different
nominal valencies which in the SIC-LSD approach are defined as: 
$N_{val}=Z-N_{core}-N_{SIC}$, where $Z$ is the atomic
number, $N_{core}$ is the number of atomic core and semicore electrons, 
and $N_{SIC}$ is the number of localized $f$ electrons. According to this definition, a
localized $f^1$ configuration of the RE ion will be referred to as trivalent in the case of Ce,
tetravalent in the case of Pr, pentavalent in the case of Nd, etc.. 

\begin{figure}
\begin{center}
\vspace{8cm}
\includegraphics[scale=0.60,angle=0]{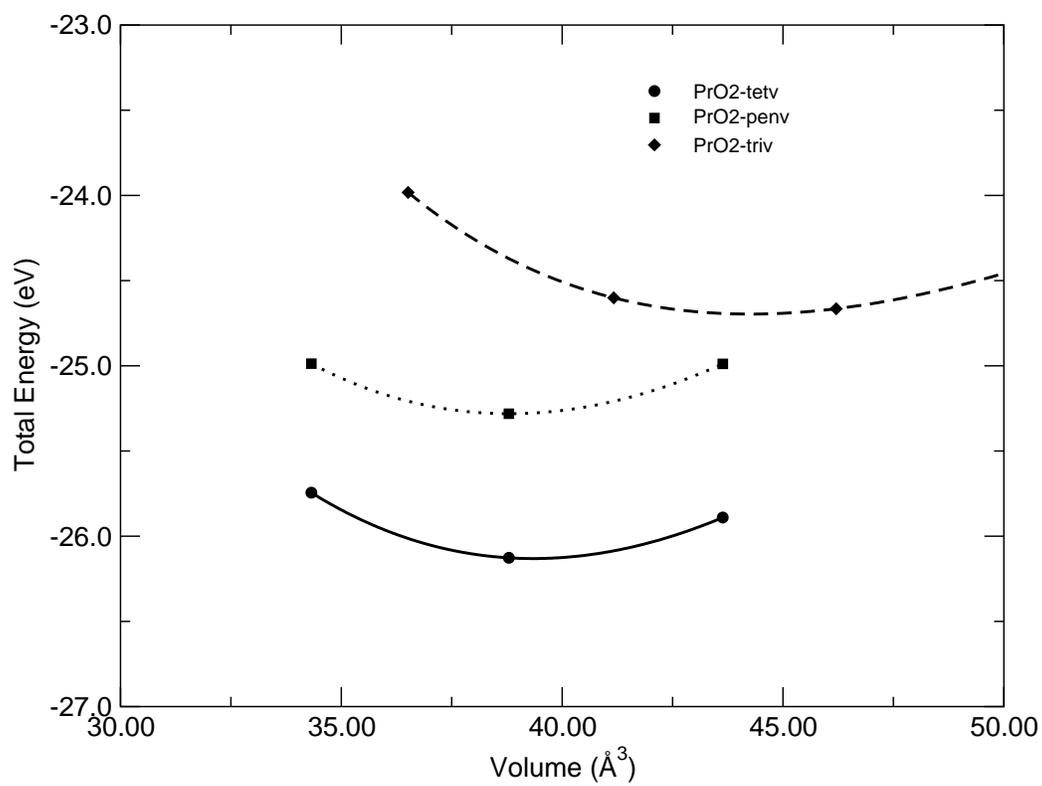}
\vspace{3cm}
\caption{
 Total energies (in eV per formula unit) as a function of volume (in \AA$^3$) for PrO$_2$,
 assuming the Pr ions to be pentavalent ($f^0$, dotted line),
tetravalent ($f^1$, solid line),
and trivalent ($f^2$, dashed line), respectively.
The global energy minimum is
obtained in the tetravalent configuration.
}
\label{energy}
\end{center}
\end{figure}

The total energies of all the RE-dioxides from CeO$_2$ to HoO$_2$ were calculated with a rare 
earth valency assumed to be either 3, 4 or 5. 
In all these cases the crystal structure was taken to be the cubic fluorite structure. 
The results are
listed in Table I, where the energy differences between the tetravalent and trivalent
configuration, $E_{IV}-E_{III}$, are displayed in column 2. Here, a negative value
indicates that the compound prefers the tetravalent groundstate configuration. The pentavalent configuration is 
found to be energetically unfavourable
in all the dioxides.
\begin{table}
\label{energies}
\begin{tabular}{|l|c|c|c|}
\hline
Compound & $E_{IV}-E_{III}$ & V$_{theo.}$&V$_{exp.}$\\ \hline
CeO$_2$ & -2.40    & 39.61 & 39.6 \\   
PrO$_2$ & -1.44  & 39.22 & 39.4 \\   
NdO$_2$ & -0.65  & 39.37 &   -  \\   
PmO$_2$ &-0.18   & 39.15 &   -  \\   
SmO$_2$ &0.49    & 42.19 &   -  \\   
EuO$_2$ &2.31    & 41.87 &   -  \\
GdO$_2$ &1.22    & 41.08 &   -  \\
TbO$_2$ &-0.27   & 36.50 & 35.6 \\
DyO$_2$ &0.05    & 39.65 &   -  \\ 
HoO$_2$ &0.46    & 39.06 &   -  \\   
\hline
\end{tabular}
\caption{
Dioxide data: Column 2: Energy difference between tetravalent and trivalent configurations (in eV).
Column 3: Theoretical volume in (\AA$^3$). 
Column 4: Experimental volume in (\AA$^3$) from Ref. \onlinecite{pearson}.
}
\end{table}
An example of the calculations involved in determining the energetically most favourable groundstate
configuration is shown in Fig.~\ref{energy} for PrO$_2$. 
In this figure, the total energy of respectively pentavalent (Pr($f^0$)), 
tetravalent (Pr($f^1$)),
and trivalent (Pr($f^2$)) PrO$_2$, is plotted as a function of volume. 
The global energy minimum is
obtained in the tetravalent configuration. 
The theoretical equilibrium volume is calculated to be $V_{tetv}$=39.2 \AA$^3$, 
which is in excellent agreement with the experimental value
$V_{exp}$=39.4 \AA$^3$.

\begin{figure}
\includegraphics[scale=0.60,angle=0]{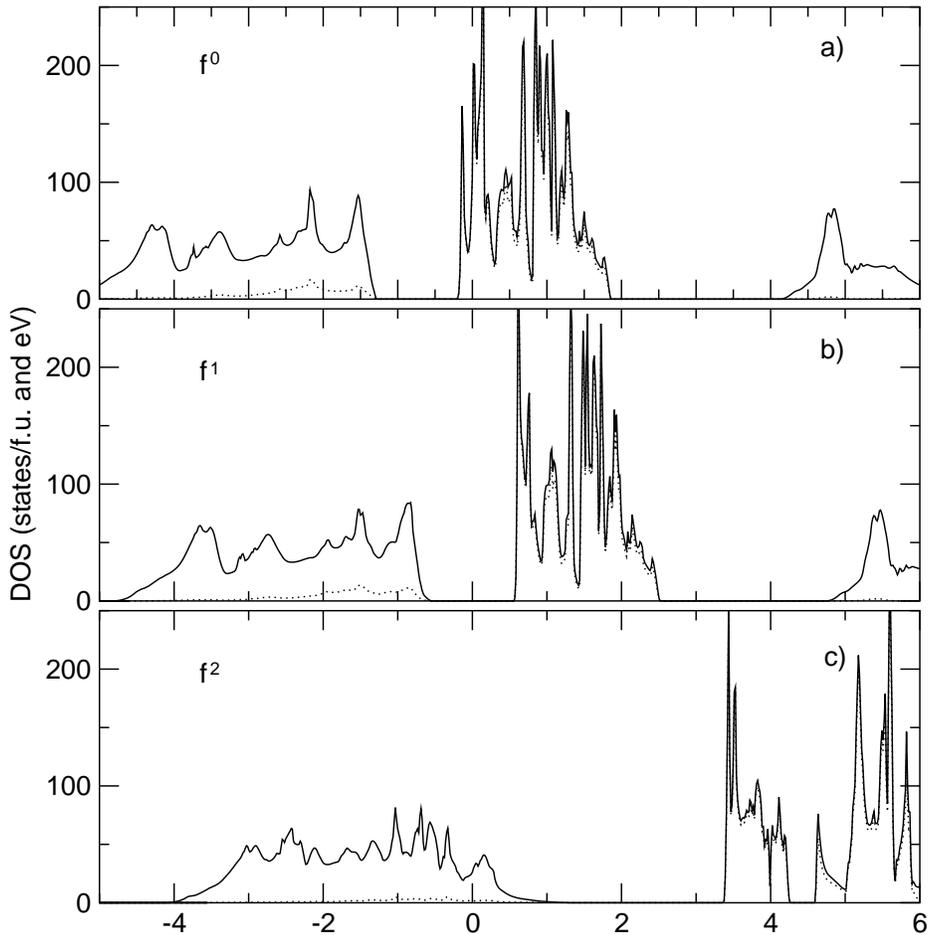}
\caption{Total DOS (solid line) and $f$-projected DOS (dotted line) for PrO$_2$, with the Pr ions in
a) the pentavalent ($f^0$) configuration b)
the tetravalent ($f^1$) configuration and c) the trivalent ($f^2$) configuration.
The energy is in units of eV, with zero marking the Fermi level.}
\label{dos}
\end{figure}

A closer examination of the electronic structure reveals why the trivalent configuration
is energetically unfavourable.
In Fig.~\ref{dos}, both the total and $f$-projected DOS of PrO$_2$ are
shown for the pentavalent (Fig. \ref{dos}a),
tetravalent (Fig. \ref{dos}b), and trivalent
(Fig. \ref{dos}c) Pr configurations.
In Fig. \ref{dos}a with all the $f$-electrons treated as delocalized,
we find the Fermi level in the $f$-peak, in accordance with the 
LSD calculations by Koelling {\it et al.},~\cite{koelling} 
and, as noted earlier, in disagreement with the observed insulating nature of
PrO$_2$. Localizing one $f$-electron, results
in the DOS of Fig. \ref{dos}b, where now the Fermi level is situated inside the large gap, 
which forms between the occupied O p states and the unoccupied conduction bands of primarily
Pr $d$ character.
Even though one $f$-electron has become localized, 
we still observe considerable $f$ hybridization with the O $p$-band, {i. e.,} the O $p$-states
have large spatial extent and their tails reach into the atomic sphere around the RE atom, where they, when 
decomposed into RE-centred spherical harmonics, attain appreciable $f$ character. 

In this respect, both localized and delocalized $f$-electrons
 coexist in the tetravalent configuration, 
which is in line with the interpretation of X-ray absorption 
studies.~\cite{karnatak,wuilloud} Finally, in the trivalent scenario of Fig. \ref{dos}c, a further
$f$-electron becomes localized, which results in the Fermi level moving down into the $p$-band. 
The O $p$-band states are depopulated to facilitate the formation of Pr $f^2$ ions, 
i.e. charge transfer is imposed on the system, 
and the associated cost in energy is significantly larger than the gain in $f$-localization energy. 
%It is remarkable how this trend also occurs in PuO$_2$ as can be seen from Fig.~3 
%in Petit {\it et al.}.~\cite{leon}
The calculated energy minimum for the trivalent $f^2$ configuration is
situated more than 1 eV above the tetravalent energy minimum (E$_{IV}$-E$_{III}$=-1.44 eV),
implying that localizing one more $f$-electron with respect to the tetravalent configuration 
results in a loss of hybridization and electrostatic energy ($\sim 2.5$ eV) that considerably exceeds the
corresponding gain in SIC localization energy ($\sim 0.95$ eV per $f$-electron). According to
the proponents of the intermediate valency, the groundstate of PrO$_2$ consists of a 50:50
mixture of 4$f^1$ and 4$f^2$ states,~\cite{bianconi,ogasawara,butorin} with an average
4$f$ occupancy of approximately 1.5 electrons. In 
the present 
picture this would imply the trivalent and tetravalent configurations to be nearly energetically
degenerate. This does not occur, as our total energy results point towards a relatively
solid tetravalent groundstate configuration for PrO$_2$. 
The calculated band gap for PrO$_2$ is approximately 1.1 eV, 
%i.e. significantly smaller than the 2.5 eV calculated by Koelling {\it et al.},~\cite{koelling} 
considerably larger than the 0.262 eV derived from conductivity measurements by 
Gardiner {\it et al.}.~\cite{gardiner} 

For CeO$_2$ we find a clearly
preferred tetravalent groundstate
configuration (the trend towards  
delocalization of the $f$-electrons is even stronger than in PrO$_2$), i.e. CeO$_2$ is best 
described in the LSD approximation, in line with results from
 earlier band structure calculations.~\cite{koelling,skorodumova} The calculated volume in this
 configuration is in good agreement with the experimental value, 
as can be seen from Table I. 

In conclusion, our calculations do not confirm an intermediate-valent groundstate for either PrO$_2$ and CeO$_2$.
Koelling {\it et al.}~\cite{koelling} argued that, the intermediate valency scenario, is related to 
an ionic description of
the RE-oxides, which can not account for the covalent $f$-$p$ bonding. 
In the ionic picture, valency is defined as the number of electrons that have been transferred from the
RE-atoms to the O-atoms, i.e. $f$-electrons do not participate in the bonding, and can only exist
as localized at the RE-sites. A given configuration thus has an integral number of $f$-electrons, and a
non-integral number of $f$-electrons can consequently only result from an intermediate-valency scenario.
In the SIC picture both localized and delocalized $f$-states coexist. The delocalized $f$-electrons
are allowed to participate in the band formation, and occur as part of the tails of O $p$ states.
The overall number of $f$-electrons is
non-integral, in analogy to the intermediate valent ionic picture, but as a consequence of the $p-f$ mixing. 

Apart from Ce and Pr, Tb also forms a dioxide, albeit under positive oxygen pressure. 
We find the tetravalent
groundstate configuration also to be energetically most favourable for TbO$_2$, 
although here the energy
difference, between the trivalent and tetravalent configurations is less pronounced 
than for 
PrO$_2$. Again the volume, calculated for the tetravalent configuration is in good agreement with the
experimental value. The remaining RE dioxides do not form in nature, and the determination of their
groundstate configuration might seem of academic interest only. 
The energy differences, E$_{IV}$-E$_{III}$, in
Table I, indicate that  NdO$_2$, and 
PmO$_2$ would also prefer the RE$^{4+}$ ion configuration. One notices however a decreasing 
affinity for the tetravalent configuration from CeO$_2$ to PmO$_2$, 
and from SmO$_2$ the energy difference
changes sign, indicating that now the trivalent configuration is energetically more favourable. 
With
increasing nuclear charge, the $f$-electrons become more tightly bound to the RE atom, 
and the decreased
overlap with neigbouring atoms results in a reduced gain in binding energy. 
Eventually it becomes more
favourable to localize an extra electron, and gain the corresponding SIC energy, 
which is what happens in
SmO$_2$, with a resulting trivalent groundstate configuration. 
In the late RE dioxides, only TbO$_2$ is
observed to be tetravalent, which is due to the fact that the Tb ion prefers a half-filled $f$-shell. 

\section{The RE-sesquioxides}

Below 2000 C, the rare-earth sesquioxides adopt three different structure types.~\cite{eyring} 
The light RE crystallize in the hexagonal La$_2$O$_3$ structure (A-type), and the heavy RE crystallize 
in the cubic Mn$_2$O$_3$ structure (C-type), also known as the bixbyite structure. 
The middle RE can be found in either the C-type structure, or the B-type, which is a monoclinic 
distortion of the C-type structure. Conversions between the different structure types are 
induced under specific temperature and pressure conditions.~\cite{hoekstra} 
We investigated the electronic structure of both the A-type and C-type sesquioxides. 
The C-type bixbyite structure was approximated by the fluorite REO$_2$ structure, 
with 1/4 of the O atoms removed. 

Starting with the hexagonal A-type structure, we see 
from Table II, column 2, that all the sesquioxides prefer the trivalent groundstate
configuration, in agreement with experiment. 
The degree of trivalency, E$_{IV}$-E$_{III}$,  increases from Ce$_2$O$_3$ to
Gd$_2$O$_3$, then decreases slightly at Tb$_2$O$_3$, to inrease again through Dy$_2$O$_3$ to Ho$_2$O$_3$. 
Similar trends were obtained for the cubic C-type sesquioxides.
The fact that the energy difference is less for Tb$_2$O$_3$ than Gd$_2$O$_3$, 
indicates that the tetravalent half-filled $f$-shell configuration becomes relatively more important. 
For this same reason, trivalent Gd$_2$O$_3$ is energetically very favourable. 

\begin{table}
\begin{tabular} {|l|c|cc|cc|}
\hline
Compound & \multicolumn{1}{c|}{$E_{IV}-E_{III}$ (eV)}
                                     & \multicolumn{2}{c|}{$V_{hexag.}$ (\AA$^3$)}&
                                       \multicolumn{2}{c|}{$V_{cubic}$(\AA$^3$)} \\
         & hexagonal
                                     & Theory & Expt.& Theory & Expt.\\
\hline
Ce$_2$O$_3$  & 0.38    & 76.4 & 79.4 & 87.27 & 87.0 \\
Pr$_2$O$_3$  & 0.78    & 75.6 & 77.5 & 85.08 & 86.7 \\
Nd$_2$O$_3$  & 0.94    & 74.0 & 76.0 & 83.52 & 85.0 \\
Pm$_2$O$_3$  & 0.97    & 72.9 & 74.5 & 82.92 & 83.0 \\
Sm$_2$O$_3$  & 1.09    & 72.0 &      & 81.27 & 81.7 \\
Eu$_2$O$_3$  & 1.13    & 70.5 &      &   -   & 80.2 \\
Gd$_2$O$_3$  & 1.29    & 68.8 &      & 79.43 & 79.0 \\
Tb$_2$O$_3$  & 1.12    & 67.6 &      & 78.21 & 77.2 \\
Dy$_2$O$_3$  & 1.21    & 66.3 &      & 77.78 & 75.9 \\
Ho$_2$O$_3$  & 1.36    & 65.2 &      & 79.72 & 74.7 \\
\hline
\end{tabular}
\caption{Sesquioxide data: Column 2: Energy difference between tetravalent and trivalent configurations
(A-type structure). The experimental $c/a$ ratio was used in the calculations.
Columns 3 and 4: A-type theoretical and experimental volumes.
Columns 5 and 6: C-type theoretical and experimental volumes. Experimental volumes are from [\onlinecite{pearson}].
}
\label{sesdata}
\end{table}

\begin{figure}
\includegraphics[scale=0.60,angle=0]{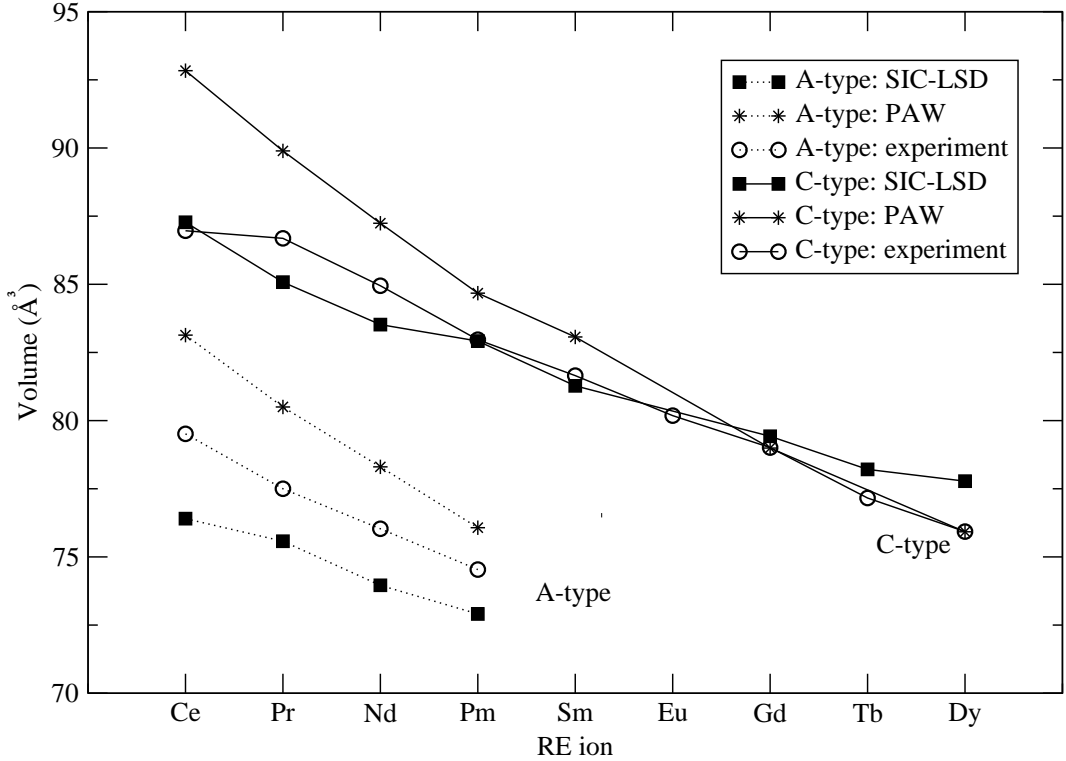}
\caption{ Calculated equilibrium volumes of the rare-earth sesquioxides,
crystallizing in the hexagonal A-type structure (dotted line), and the cubic C-type structure (full line).
The stars, circles, and squares refer to the PAW,\cite{hirosaki} SIC-LSD (present), and experimental
results.\cite{pearson}
}
\label{latt}
\end{figure}

Apart from the extraordinary stability of the  half-filled shell configuration, 
the general tendency towards trivalency is clearly related to the increasing 
localization of the $f$-electrons with atomic number. The highly directional $f$-orbitals 
are only partially able to screen each other from the attractive force of the nucleus, 
which results in a steadily increasing effective nuclear charge with increasing number of $f$-electrons. 
The increase in localization leads to the well known lanthanide contraction, i.e. 
the decrease in ionic radii across the rare earth series, which is also reflected in the lattice 
parameters of the  sesquioxides. 
In Table II, the experimental and theoretical volumes are compared, for the A-type (columns 3 and 4) 
and C-type (columns 5 and 6) sesquioxides. The data are also
illustrated in Fig. 3. Good agreement is seen with respect to both the overall trends and 
the absolute values. 
For the A-type hexagonal structures, the experimental $c/a$ ratio was used, {\it i.e.,} no optimization
of this parameter was attempted. 
From Fig. \ref{latt}, we notice that the agreement between theory and 
experiment is considerably better for the C-type structure than for the A-type one. 
This might be related to the ASA used in the calculations. This geometrical approximation is likely 
less reliable when applied to the hexagonal A-structures, 
than when applied to the higher symmetry cubic C-structure. 
Figure \ref{latt} also includes the equilibrium volumes calculateded by Hirosaki {\it et al.}~\cite{hirosaki}, 
using the projector augmented-wave (PAW) pseudopotential  method, with the localized partly filled $f$-shell 
being treated as part of the core. 
%The agreement between calculated and observed equilibrium volumes
%are in all cases excellent, worst case being $\sim 3$ \% deviation.
For the A-type structure, 
the SIC-LSD calculations consistently yield lower volumes than observed, while the PAW values
are consistently above.
For the C-type structure, the present SIC-LSD calculations agree somewhat better with observations than the 
PAW results, in particular for the earlier REs, which may be due to the core approximation
made for the $f$-electrons in the work of Ref. [\onlinecite{hirosaki}], and which may be too restrictive
in the earlier RE sesquioxides. 

\begin{figure}

\includegraphics[scale=0.60,angle=0]{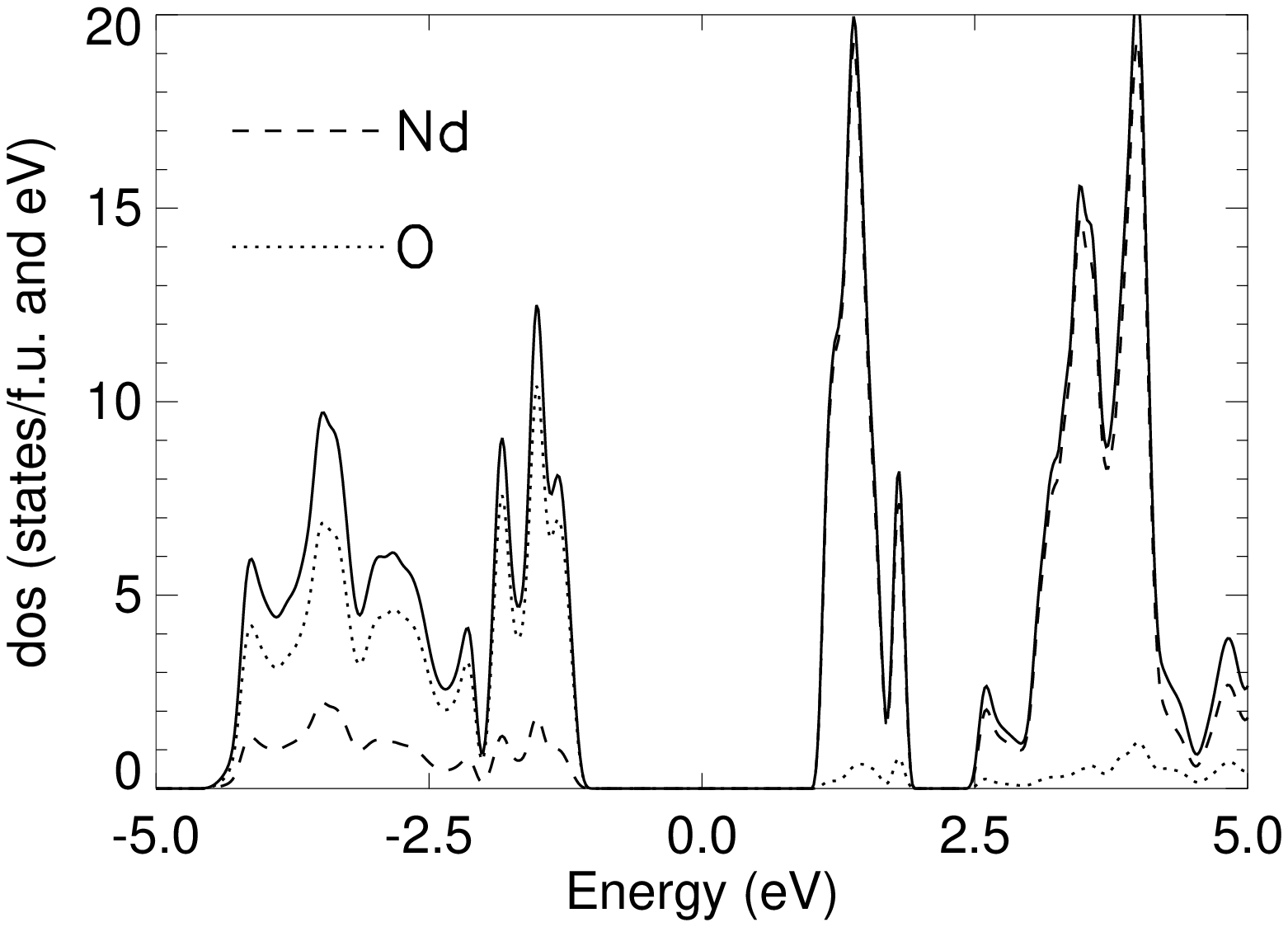}
\caption{Total DOS of Nd$_2$O$_3$ in the trivalent ($f^3$ )configuration in the A-type hexagonal structure
at the calculated equilibrium volume. Energy is in eV, with 0 marking the mid-gap position, while DOS is in
states per formula units and eV.}
\label{Nd2O3dos}

\end{figure}

\begin{figure}

\includegraphics[scale=0.60,angle=0]{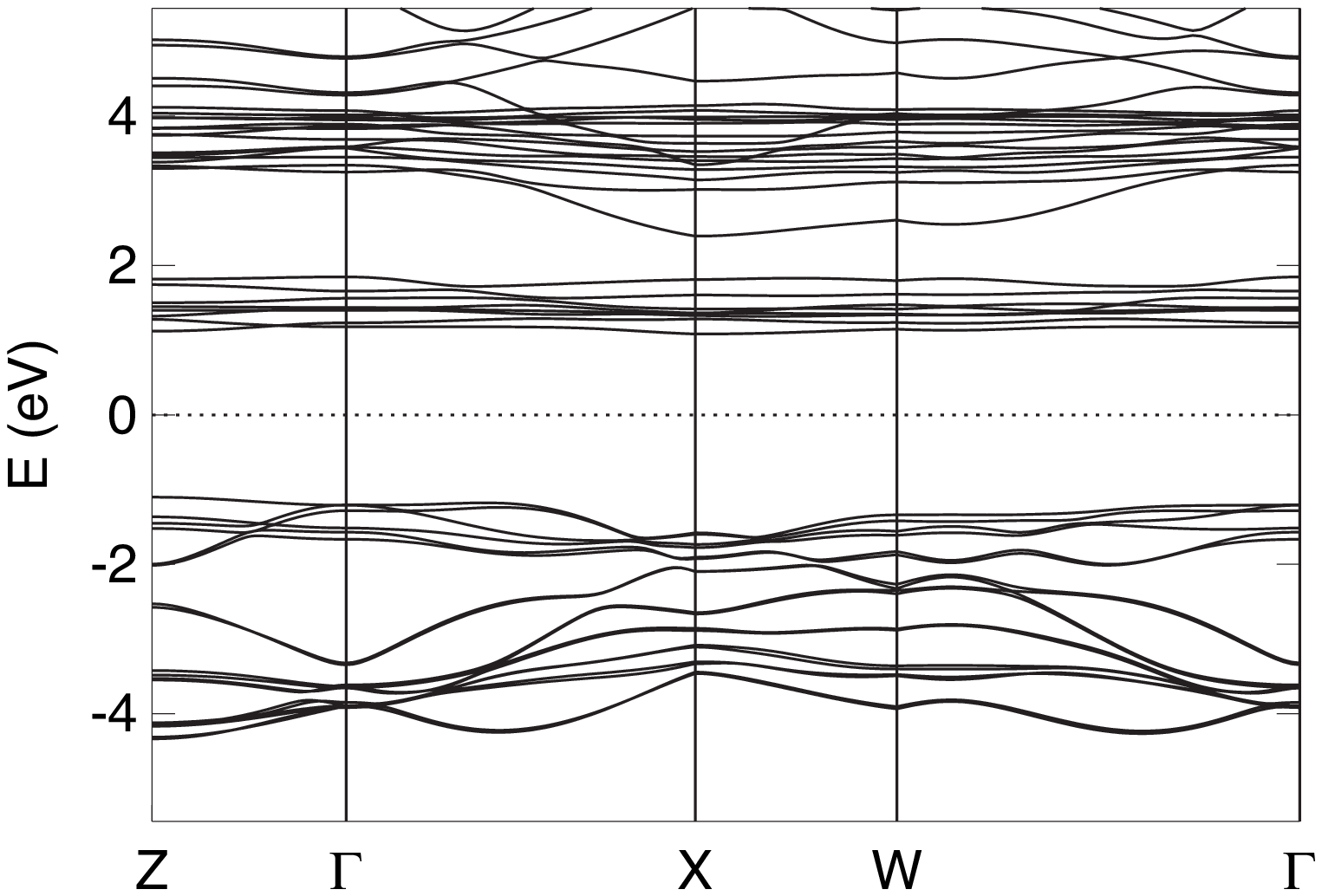}
\caption{Bandstructure of A-type Nd$_2$O$_3$ in the trivalent configuration.}
\label{Nd2O3bs}

\end{figure}

The DOS and bandstructures for all the sesquioxides are quite
similar and as a representative example in the Figs. \ref{Nd2O3dos} and \ref{Nd2O3bs}
we present those  
of Nd$_2$O$_3$, as calculated in the trivalent 
ground state 4$f^3$ configuration. 
The broad band below the Fermi level originates from the O-$p$ states.   
Hybridization and charge transfer result in this band being completely filled, leaving 
behind an empty $f$-peak situated in the gap between the valence and (non-$f$)conduction bands. 
We find all the sesquioxides to be insulators.

\begin{figure}
\includegraphics[scale=0.60,angle=0]{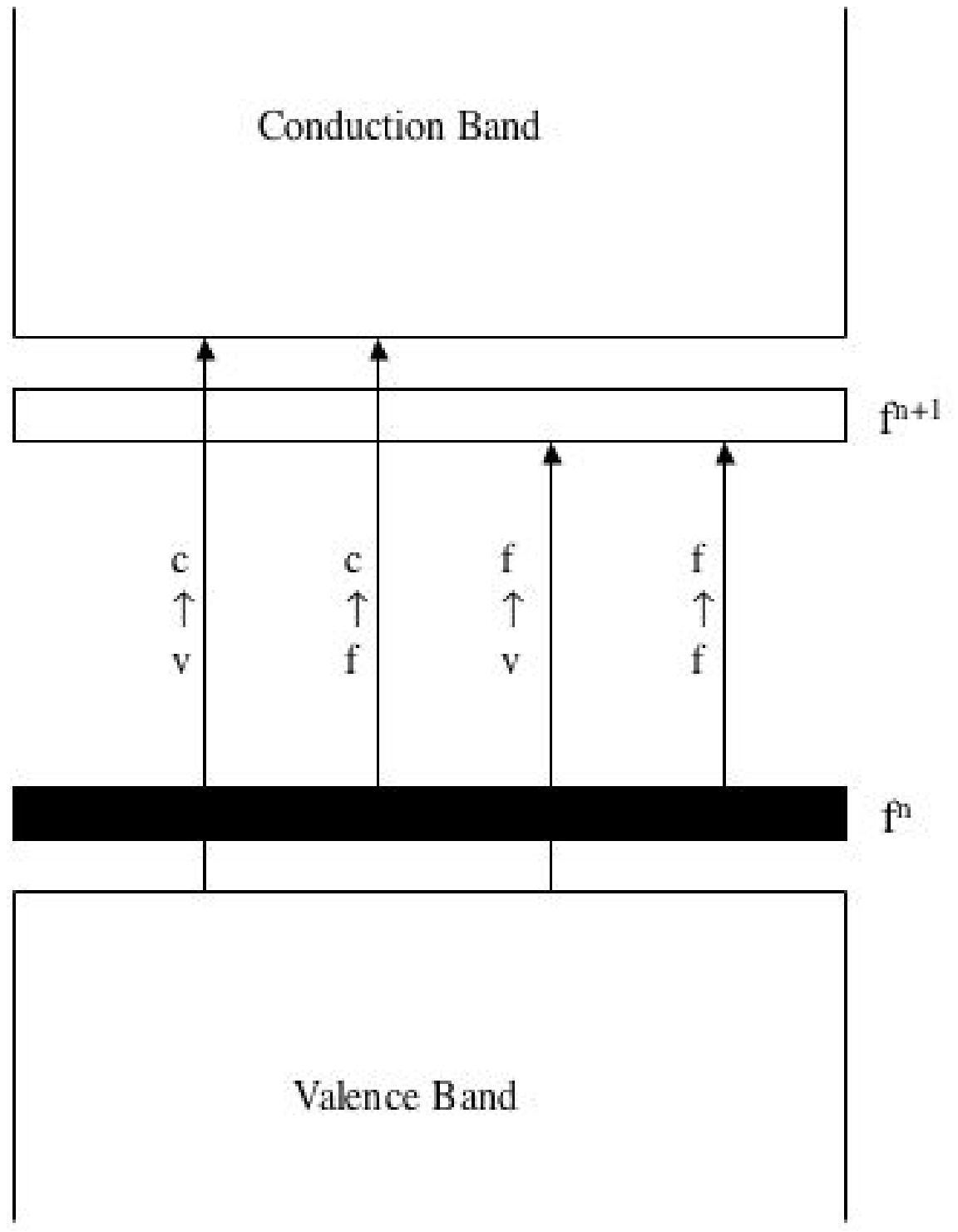}
\caption{Schematic representation of the RE sesquioxide band gap. The $f^n$ and $f^{n+1}$ depict respectively
the occupied and empty $f$-bands.}.
\label{cbmvbm}
\end{figure}

In Table \ref{sesqui}, we compare the calculated band gaps for the A-type structure, to the experimental values, 
as obtained respectively from optical~\cite{prokofiev} (column 4) and conductivity~\cite{lal,subba} 
(columns 5 and 6) measurements. The two sets of
conductivity measurements refer respectively to temperature regions above~\cite{lal} (column 5) and
below~\cite{subba} (column 6) T=800 K. We notice a considerable 
discrepancy between the optical and conductivity
measurements, and the two sets of results have been interpreted slightly differently. 
In Fig. \ref{cbmvbm} we show schematically the various transitions that can be envisaged to occur. The actual
transition depends crucially on the position of the occupied and empty $f$-states, respectively $f^n$ and $f^{n+1}$. The
interpretation of the optical data is that the empty $f$-levels are situated above the conduction band minimum (CBM),
and the energy gap E$_g$ is entirely determined by the gap between the occupied $f$- and the CBM. If the $f$-states
move below the valence band maximum (VBM), the band gap becomes largest (i.e. $v\rightarrow c$). The
activation energies E$_a$ in column 5, obtained from conductivity measurements (using conductivity$ \sim  e^{-E_a/2kT}$)
indicate that, for some compounds, the transitions is from ($v\rightarrow f$), i.e. that the empty levels can be situated
in the gap beteen VBM and CBM. 

One could speculate whether the absence of ($v\rightarrow f$) transitions in the optical spectrum 
is due to this transition being optically forbidden, whilst no such
constraint exists for the thermal excitations, which go into the unoccupied $f$-bands. 
However this can not explain
the relatively large difference in the gap energy, observed for example in the case of Gd$_2$O$_3$,
where both optical and conductivity mesurements agree on the nature of the gap being ($v\rightarrow c$).  
From the theoretical data we see that the gap between valence and
conduction bands ($v\rightarrow c$) shows a rather slow increase
across the RE series, whereas the 
gap between valence and $f$-band ($v\rightarrow f$)
decreases from Ce$_2$O$_3$ to Eu$_2$O$_3$, 
and from Gd$_2$O$_3$ to Dy$_2$O$_3$. For Ce$_2$O$_3$, Pr$_2$O$_3$
and Gd$_2$O$_3$ no separate edge for the non-f conduction bands was found in the
calculations.
The direct comparison between
calculated and experimental gaps is tricky. The SIC-LSD approach cannot accurately determine the removal
energies of localized states, and the bare $f$-bands always appear
at too high binding energies due to the neglect of screening
and relaxation effects.~\cite{temmerman} With the SI corrected $f$-states situated below the valence band
edge, the calculated band-gaps are of either $v$-$f$ or $v$-$c$ character.
The observed trend can be explained by the fact
that with increasing atomic number the $f$-orbitals become increasingly localized, and the energy of the $f$-shell
decreases. At Gd$_2$O$_3$ all seven majority spins are SI-corrected, and the minority states are situated above the 
conduction band minimum. The $v\rightarrow f$ gap starts decreasing again from Tb$_2$O$_3$ as the miniority states 
are pulled into the gap between valence and conduction band.

\begin{table}
\label{sesqui}
\begin{tabular} {|l|cc|ccc|}
\hline
Compound &  \multicolumn{5}{c|}{$E_{gap}$ (eV)}\\
         & \multicolumn{2}{c|}{Theory} & &Experiment&  \\
%        & hexagonal & \multicolumn{1}{c|}{cubic} &optical& $\sigma$ (T=800-1200 K) & $\sigma$ (T$<$ 800 K)\\
         & $v\rightarrow f$& \multicolumn{1}{c|}{$v\rightarrow c$} &optical& $\sigma$ (T=800-1200 K) & $\sigma$ (T$<$ 800 K)\\
\hline
Ce$_2$O$_3$     & 3.20 & 3.20 & 2.4& -&- \\
Pr$_2$O$_3$     & 2.62 & 2.62 & 4.6& -&0.8$^C$ \\
Nd$_2$O$_3$     & 2.13 & 3.58 & 4.4& 2.36$^A$ ($f\rightarrow c$)&- \\
Pm$_2$O$_3$     & 1.66 & 3.77 &  - & -&- \\
Sm$_2$O$_3$     & 1.04 & 3.80 & 5.0& 2.12$^B$ ($v\rightarrow f$)&1.2$^C$ \\
Eu$_2$O$_3$     & 0.63 & 3.95 & 4.4& 1.84$^C$ ($v\rightarrow f$)&1.2$^C$  \\
Gd$_2$O$_3$     & 3.13 & 3.13 & 5.3& 2.64$^C$ ($v\rightarrow c$)&1.0$^C$ \\
Tb$_2$O$_3$     & 2.27 & 4.32 & 4.1& 2.24$^B$ ($f\rightarrow c$)&0.8$^C$  \\
Dy$_2$O$_3$     & 1.71 & 4.39 & 4.9& 2.82$^C$ ($v\rightarrow c$)&- \\
Ho$_2$O$_3$     & 1.26 & 4.48 & 5.2& 2.74$^C$ ($v\rightarrow c$)&1.4$^C$ \\
\hline
\end{tabular}
\caption{Sesquioxide band-gaps. Theoretical values for A-type structure: valence to $f$-band (column 2), 
valence to conduction band (column 3). 
Experimental values: optical measurements (column 4), columns 5 and 6 from conductivity measurements above 
and below 800 K respectively.
}
\end{table}
Our calculated values for
$v\rightarrow c$ are
consistently smaller than the optical E$_g$.
If the optical E$_g$'s turn out 
to be a better representation of the actual band gaps, the reason for the difference with our calculated 
results could be twofold:  
The LSD is known to underestimate gaps between occupied and unoccupied bands, 
and in addition, the unoccupied $f$ bands may be positioned extraordinarily low due to the neglect of
correlation effects beyond the LSD for these. In order to discuss in more detail the accuracy of the calculated 
energy gaps, one would need to further investigate the reason behind the considerable discrepancy between 
the optical and conductivity measurements.

\section{Oxidation}

So far, the present theory has demonstrated that all the sesquioxides prefer the trivalent groundstate configuration, 
whilst the dioxides can be separated into tetravalent light RE-dioxides, and trivalent heavy RE-dioxides, 
with the exception of TbO$_2$, which also prefers the tetravalent groundstate. 
From experiment however we know that all the RE-sesquioxides are found in nature, 
whilst the only dioxides that occur naturally are CeO$_2$, PrO$_2$ and TbO$_2$. 
With respect to the valencies, these two sets of information lead us to conclude that the 
RE-sesquioxides are trivalent, and the naturally occurring RE-dioxides are tetravalent,
{\it i. e.,} 
 the oxidation process from sesquioxide to dioxide 
 goes hand in hand with the delocalization of an extra $f$-electron.
The extraordinary catalytic properties of Ceria are according 
to Skorodumova {\it et al.}~\cite{skorodumova1} based on the coupling 
between $f$-electron localization and oxidation. 
Recent calculations~\cite{leon} have similarly shown that, the possible existence of the higher oxidation state, 
PuO$_{2+x}$,~\cite{Haschke} can be explained by the occurence of a Pu pentavalent state, which is reduced to a 
tetravalent state in PuO$_2$.
The question remains however, why the dioxide only exists for Ce, Pr, and Tb. 
For example, from the data in Table I it follows that Nd in NdO$_2$ also readily takes
the tetravalent configuration, but nevertheless this compound does not occur naturally.

To discuss this issue, we consider the oxidation process:
\begin{eqnarray}
\mbox{RE}_2\mbox{O}_3+\mbox{$\frac{1}{2}$O}_2 &\rightleftharpoons &2\mbox{REO}_2. 
\label{redox}
\end{eqnarray}
The balance of this reaction in general will depend on the Gibb's free energy of
the reactants at the given temperature and pressure. The ab-initio calculations of these quantities
are beyond the capability of the present
theory. However we can still to some extent analyze the reaction~(\ref{redox}) by looking at the 
zero temperature and zero pressure limit. In that case the free energy difference 
between the reactants reduces to
the corresponding total energy difference, as obtained by the SIC calculations:
\begin{eqnarray}
E_{ox}\equiv 2E^{SIC}(\mbox{REO}_2)-E^{SIC}(\mbox{RE}_2\mbox{O}_3)-\mu_O.
\label{redox-E}
\end{eqnarray}
Still, the comparison of the dioxide and sesquioxide
total energies is not straightforward due to the 
approximations of the present SIC-LSD implementation, most
notably the ASA.
The total energies depend on the ASA radii chosen
for the constituent atoms. 
 The problem is most severe for the low symmetry hexagonal structure of the sesquioxides, 
and is the most probable reason why the calculated equilibrium volumes of the 
hexagonal sesquioxides are in somewhat poorer agreement with experiment, cf. Table II.  
Unfortunately, the SIC-LSD approach has only been implemented
within the tight-binding LMTO-ASA method and not in its full
potential version FP-LMTO, where the problems
associated with the ASA would not occur. However, in the $f^0$ groundstate configuration,
the relative error introduced by the ASA, can be estimated from a comparision of the total energies, 
as obtained respectively with a FP-LMTO method and our LMTO-ASA
implementation of the SIC-LSD method.  
This provides a FP-ASA total energy correction,
which can be used to calibrate the corresponding energies calculated with the SIC-LSD. 
Since the energy differences between the valency configurations
of a given compound are almost unaffected by the chosen ASA radii, 
for a given compound the same correction
can be applied to the total energies in the different localization configurations. 

\begin{figure}
\vspace{3cm}
\includegraphics[scale=0.60,angle=0]{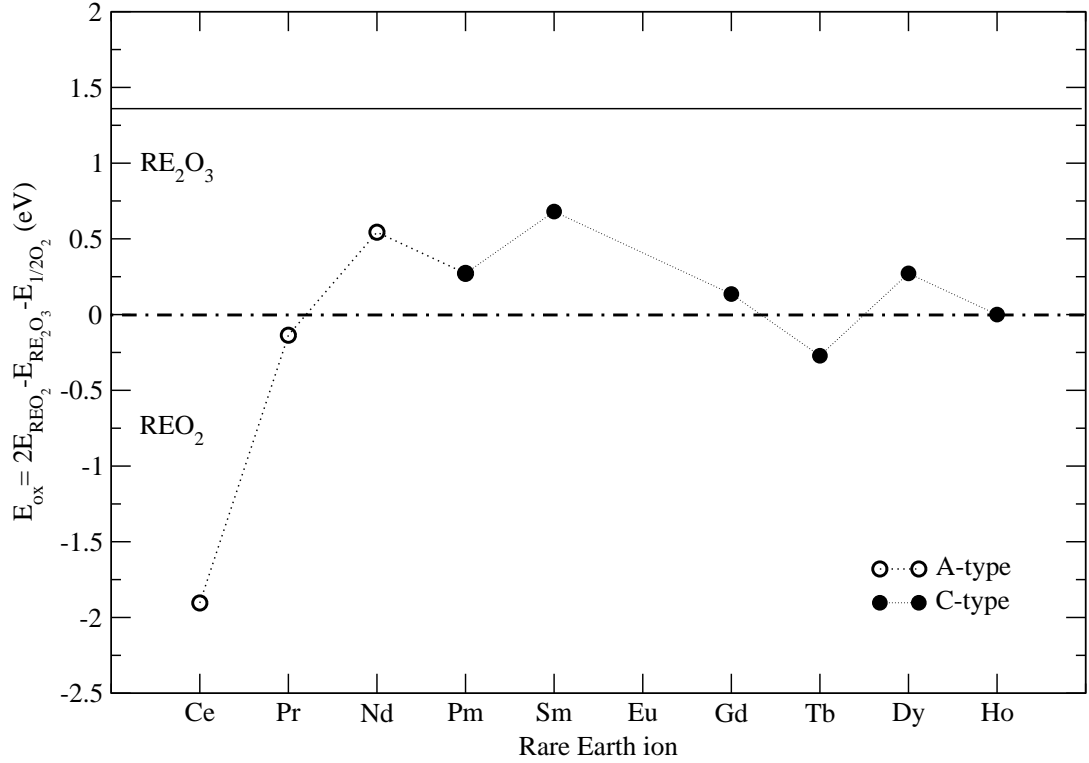}
\vspace{3cm}
\caption{Oxidation energies, $E_{ox}$ in Eq. (\ref{redox-E}), for the rare earths Ce to Ho. 
The full circle is for the cubic C-type sesquioxide, the empty circle for the A-type hexagonal structure
of the sesquioxide.
Negative values indicate that the dioxide is stable. The chemical potential of free O is taken as 
$\mu_O=-6.12$ eV. The solid line just below 1.5 eV indicates the dioxide/sesquioxide borderline in case the FP-LMTO calulated value $\mu_O=-4.76$ eV is used.  
}
\label{oxidat}
\end{figure}

In the following, we assume that the dioxides and the C-type sesquioxides, which both crystallize in a cubic structure, 
are adquately described within the ASA approximation. The FP-ASA correction is therefore only applied to the A-type 
sesquioxide, which crystallizes in the hexagonal structure. In Table \ref{oxidattable} the resulting oxidation energies~(\ref{redox-E}) 
assuming either the A or C-type sesquioxides are given in columns 2 and 3 respectively. 
The chemical potential used, is $\mu_O=-6.12$ eV (relative to free atoms). A negative/positive energy balance means that 
the formation of the dioxide/sesquioxide is energetically most favourable. 
\begin{table}
\begin{tabular}{|l|c|c|}  
\hline
REO$_2$/RE$_2$O$_3$&E$_{ox}$(A-RE$_2$O$_3$) &E$_{ox}$(C-RE$_2$O$_3$)\\
\hline
CeO$_2$/Ce$_2$O$_3$  & {\bf -1.90} & -3.54 \\
\hline
PrO$_2$/Pr$_2$O$_3$&  {\bf -0.14} & -1.90  \\ 
\hline 
NdO$_2$/Nd$_2$O$_3$  &  {\bf 0.54} &  -0.54 \\
\hline
PmO$_2$/Pm$_2$O$_3$  &  0.00 &  {\bf  0.27} \\
\hline
SmO$_2$/Sm$_2$O$_3$ &   -0.82 & {\bf 0.68}  \\
\hline
GdO$_2$/Gd$_2$O$_3$ &  -10.88 & {\bf  0.14}  \\
\hline
TbO$_2$/Tb$_2$O$_3$  &  -8.16 & {\bf -0.27}  \\
\hline
DyO$_2$/Dy$_2$O$_3$  & -4.08 & {\bf  0.27}  \\
\hline
HoO$_2$/Ho$_2$O$_3$  &  -0.68 &  {\bf  0.00}  \\
\hline                                 
\end{tabular}                        
\label{oxidattable}
\caption{Oxidation energies in eV (from equation (2)). Column 2: assuming the 
oxidation process involves the A-type sesquioxide, Column 3: assuming the oxidation
process involves the C-type sesquioxide. Negative/positive energies  indicate that
the dioxide/sesquioxide  is energetically most favourable. The bold characters indicate
which type of sesquioxide will preferentially be formed, if at all.}
\end{table}
From the table we see that in the case of Ce, Pr, and Tb, the dioxide is preferred energetically with respect to
both the A- and C-type sesquioxide. We
also find that for Ce and Pr the A-type sesquioxide is least unfavourable with respect to the dioxide (as indicated
by the relatively smaller oxidation energy), whilst for
Tb it is the C-type sesquioxide that is closest in energy to the dioxide. This is relevant since it indicates
which sesquioxide structure will form in the reduction process.
For Nd, we find the dioxide to be more favourable than the C-type sesquoxide, but less favourable than the
A-type sesquioxide, i.e. at T=0, Nd exists as A-type Nd$_2$O$_3$. NdO2 may exist as a metastable phase, since
steric effects may hinder the reduction of this compound into the
hexagonal sesquioxide. As discussed, up to now the synthesis of NdO2
has not been successful. For the remaining RE, the C-type sesquioxide
is more stable than the dioxide, as indicated by the positive oxidation energies.  
The relevant oxidation energies (indicated by boldface in Table \ref{oxidattable}) are plotted in Fig.~\ref{oxidat}
as a function of rare earth ion. The overall picture is that Ce, Pr, and Tb exist as dioxides, whilst the remaining 
RE prefer the sesquioxide, in agreement with experiment. It also shows that with respect to the sesquioxides, 
the later RE from Pm onwards crystallize in the C-type structure, whilst reduction of the early REO$_2$ will result 
in sesquioxide crystallizing in the A-type, which again would be in good agreement with experiment.

There is some uncertainty in our results associated with the value of the oxygen chemical potential. 
We have conveniently used a value of $\mu_O=-6.12$ eV, as it allows us to indicate the
trend in oxidation energies more clearly. 
Actually, using a FP-LMTO calculation, with the O$_2$ molecule put into a lattice with large lattice constant, we 
find $\mu_O=-4.76$ eV, i.e. 1.36 eV less than the value used in Table \ref{oxidattable}. Using this value in
Fig. \ref{oxidat} gives a qualitatively slightly different picture, as the oxide/sesquioxide borderline is
now situated 1.36 eV higher (solid line), which means that in this scenario, the dioxide is always preferred. 
Again other LSD total energy calculations give  a binding energy of 7.48 eV for the O$_2$ molecule, and the
experimental value is 5.17 eV.~\cite{jones89}.
We should also notice here, that the calculated total 
energies  apply to $ T= 0$, and
under actual conditions, the reduction (left side of equation (1)) will be favoured by the higher entropy, 
as it leads to release of O$_2$. Higher temperature thus will work in favour of sesquioxide formation.  
It seems that the existence of tetravalent CeO$_2$, PrO$_2$ and TbO$_2$ is an indication that their sesquioxides 
will readily oxidize. Conversely, we could interpret the trivalent configuration of the other 
RE-dioxides as evidence that their sesquioxides will not further oxidize.

\section{Conclusion}
In conclusion, we used the SIC-LSD method to analyze the valencies of the rare-earth dioxides and
sesquioxides. 
We find from total energy considerations, that CeO$_2$, PrO$_2$, and TbO$_2$ prefer the tetravalent configuration, 
with respectively zero, one, and seven localized $f$-electrons. The intermediate valency scenario is not found to 
be energetically viable, but non-integral total $f$-occupancies do occur due to covalent
$p-f$ mixing. The sesquioxides are all found to prefer the trivalent
configuration, leading to insulating groundstates in agreement with experiment. 
The calculated equilibrium volumes for both the 
dioxides and the sesquioxides are in good agreement with the experimental values. 
The oxidation reaction from sesquioxide to dioxide is found to be accompanied by $f$-electron delocalization, 
and consequently only occurs for the early lanthanides and Tb, for which the $f$-electrons are 
least tightly bound. 

\begin{acknowledgments}
This work has been partially funded by the Training and Mobility Network on
'Electronic Structure Calculation of Materials Properties and Processes for 
Industry and Basic Sciences' (contract:FMRX-CT98-0178) and by the Research Training
Network on 'Ab-initio Computation of Electronic Properties of f-electron Materials'
(contract:HPRN-CT-2002-00295). Work of L.P. was sponsored in part 
by the Office of Basic Energy Sciences, U.S. Department of Energy.
The Oak Ridge National Laboratory is managed by UT-Battelle LLC for the 
Department of Energy under Contract
No. DE-AC05-00OR22725.

\end{acknowledgments}

\end{document}